\documentclass{aastex62}

\received{...}
\revised{...}
\accepted{...}

\submitjournal{ApJ}

\shorttitle{Solar eruptions cause heating of nearby quiescent loops}
\shortauthors{Li et al.}

\begin{document}

\title{Heating of quiescent coronal loops caused by nearby eruptions observed with the Solar Dynamics Observatory and the Solar Upper Transition Region Imager}

\correspondingauthor{Leping Li}
\email{lepingli@nao.cas.cn}

\author[0000-0001-5776-056X]{Leping Li}
\affil{National Astronomical Observatories, Chinese Academy of Sciences, Beijing 100101, Peoples's Republic of China}
\affiliation{University of Chinese Academy of Sciences, Beijing 100049, Peoples's Republic of China}
\affiliation{Key Laboratory of Solar Activity and Space Weather, National Space Science Center, Chinese Academy of Sciences, Beijing 100190, People's Republic of China}

\author[0000-0002-1369-1758]{Hui Tian}
\affil{School of Earth and Space Sciences, Peking University, Beijing 100871, People's Republic of China}
\affiliation{Key Laboratory of Solar Activity and Space Weather, National Space Science Center, Chinese Academy of Sciences, Beijing 100190, People's Republic of China}
\affil{National Astronomical Observatories, Chinese Academy of Sciences, Beijing 100101, Peoples's Republic of China}

\author[0000-0001-6076-9370]{Huadong Chen}
\affil{National Astronomical Observatories, Chinese Academy of Sciences, Beijing 100101, Peoples's Republic of China}
\affiliation{University of Chinese Academy of Sciences, Beijing 100049, Peoples's Republic of China}
\affiliation{Key Laboratory of Solar Activity and Space Weather, National Space Science Center, Chinese Academy of Sciences, Beijing 100190, People's Republic of China}

\author[0000-0001-5705-661X]{Hongqiang Song}
\affiliation{Shandong Provincial Key Laboratory of Optical Astronomy and Solar-Terrestrial Environment, and Institute of Space Sciences, Shandong University, Weihai, Shandong 264209, Peoples's Republic of China}

\author[0000-0003-4804-5673]{Zhenyong Hou}
\affil{School of Earth and Space Sciences, Peking University, Beijing 100871, People's Republic of China}

\author[0000-0003-2686-9153]{Xianyong Bai}
\affil{National Astronomical Observatories, Chinese Academy of Sciences, Beijing 100101, Peoples's Republic of China}
\affiliation{University of Chinese Academy of Sciences, Beijing 100049, Peoples's Republic of China}
\affiliation{Key Laboratory of Solar Activity and Space Weather, National Space Science Center, Chinese Academy of Sciences, Beijing 100190, People's Republic of China}

\author[0000-0001-8950-3875]{Kaifan Ji}
\affil{Yunnan Observatories, Chinese Academy of Sciences, Kunming 650216, People's Republic of China}

\author{Yuanyong Deng}
\affil{National Astronomical Observatories, Chinese Academy of Sciences, Beijing 100101, Peoples's Republic of China}
\affiliation{University of Chinese Academy of Sciences, Beijing 100049, Peoples's Republic of China}
\affiliation{Key Laboratory of Solar Activity and Space Weather, National Space Science Center, Chinese Academy of Sciences, Beijing 100190, People's Republic of China}

\begin{abstract}

How structures, e.g., magnetic loops, in the upper atmosphere, i.e., the transition region and corona, are heated and sustained is one of the major unresolved issues in solar and stellar physics.
Various theoretical and observational studies on the heating of coronal loops have been undertaken.
The heating of quiescent loops caused by eruptions is, however, rarely observed.
In this study, employing data from the Solar Dynamics Observatory (SDO) and Solar Upper Transition Region Imager (SUTRI), we report the heating of quiescent loops associated with nearby eruptions.
In active regions (ARs) 13092 and 13093, a long filament and a short filament, and their overlying loops are observed on 2022 September 4.
In AR 13093, a warm channel erupted toward the northeast, whose material moved along its axis toward the northwest under the long filament, turned to the west above the long filament, and divided into two branches falling to the solar surface.
Subsequently, the short filament erupted toward the southeast.
Associated with these two eruptions, the quiescent loops overlying the long filament appeared in SDO/Atmospheric Imaging Assembly (AIA) high-temperature images, indicating the heating of loops.
During the heating, signature of magnetic reconnection between loops is identified, including the inflowing motions of loops, and the formation of X-type structures and newly reconnected loops.
The heated loops then cooled down.
They appeared sequentially in AIA and SUTRI lower-temperature images.
All the results suggest that the quiescent loops are heated by reconnection between loops caused by the nearby warm channel and filament eruptions.

\end{abstract}


\section{Introduction} \label{sec:int}

In solar and stellar astrophysics, how structures in the upper atmosphere, i.e., the transition region and corona, are heated and sustained is one of the major unresolved issues \citep[see a review in][and references therein]{2006SoPh..234...41K}.
As the main building blocks of the solar corona \citep{2013A&A...556A.104P, 2014A&A...570A..93L, 2017A&A...599A.137B}, coronal loops are considered to be the ideal features to investigate the dominant heating mechanism(s) in the upper solar atmosphere \citep{2000ApJ...541.1059A, 2004ApJ...601..530S, 2022FrASS...920116A}.
Depending on their temperatures, coronal loops are usually classified into two types: warm ($\sim$1\,MK) and hot ($>$2\,MK) loops \citep{2003ApJ...590..547A, 2009ApJ...695..642U}.
They are prominently seen in extreme ultraviolet (EUV) and X-rays \citep{2009SSRv..149...65D, 2014LRSP...11....4R}.
For understanding the coronal heating, many studies of coronal loops have been done in theory and observations \citep[e.g.,][]{2006ChJAA...6..608F, 2006SoPh..234...41K, 2021ApJ...915...39H}.

In non-flaring conditions, current scenarios of coronal heating rely on either the dissipation of magnetohydrodynamic (MHD) waves \citep[e.g.,][]{1961ApJ...134..347O, 2011ApJ...736....3V, 2020SSRv..216..140V}, or the Ohmic dissipation of current sheets in nanoflares \citep[e.g.,][]{1988ApJ...330..474P, 1991SoPh..133..357H, 2008ApJ...688..669A}.
In the solar atmosphere, MHD waves are ubiquitously observed \citep[e.g.,][]{2012ApJ...759..144T, 2020Sci...369..694Y}.
They can heat the quiet corona, e.g., the corona outside the active regions \citep[ARs;][]{2007Sci...317.1192T, 2011Natur.475..477M, 2023ApJ...942L...2Z}. 
However, the observed wave power is generally two-orders-of-magnitude too weak to power the AR corona \citep[see reviews in][and references therein]{1977ARA&A..15..363W, 2015RSPTA.37340261A}.
In the nanoflare heating models, the sudden heating results in the abrupt temperature increase that leads to an intense heat flux into the transition region and chromosphere \citep{1997ApJ...478..799C}.
This drives an evaporative upflow, i.e., the chromospheric evaporation, into the coronal loops \citep{2006ApJ...647.1452P}.

By comparing the observed and theoretical emission measure (EM) distributions in an AR core, \citet{2010ApJ...723..713T} suggested that the hot loops are heated by nanoflares.
By employing the EUV images taken by the Atmospheric Imaging Assembly \citep[AIA;][]{2012SoPh..275...17L} on board the Solar Dynamic Observatory \citep[SDO;][]{2012SoPh..275....3P}, \citet{2012ApJ...753...35V} analyzed the light curves of coronal loops.
By comparing the observations with theoretical models, they proposed that the loops both in and surrounding the AR cores are heated by impulsive nanoflares.
By measuring the Doppler shifts in AR moss in the lower-temperature lines, e.g., Ne VIII and C IV, \citet{2013ApJ...767..107W} indicated that the hot loops are impulsively heated.
By using AIA multi-wavelength images, \citet{2015A&A...583A.109L} analyzed a fast heating of AR loops that go along with evaporation of material.
By comparing the observations with the predictions from a one-dimensional loop nanoflare model \citep{2006ApJ...647.1452P}, they proposed that the observations support the nanoflare process in (almost) all aspects.
Recently, \citet{2017ApJS..229....4C, 2018A&A...615L...9C} suggested that magnetic flux cancellation and reconnection at the base of coronal loops due to mixed polarity fields can play an important role in energizing AR loops driven by nanoflares.

Once the heating stops, the heated loops cool down.
For coronal loops, the light curves of channels that observe plasma with lower temperatures reach their peaks progressively at later times than channels that sample plasma with higher temperatures \citep{2001SoPh..198..325S, 2004A&A...424..289M, 2012A&A...537A.152P}.
This time lag has been interpreted as the result of hot coronal plasma cooling down.
By employing the EUV Imaging Spectrometer (EIS) data on board the Hinode, \citet{2009ApJ...695..642U} indicated that coronal loops cool down to transition region temperatures after a heat deposition.
\citet{2012ApJ...753...35V} found a time lag, i.e., the cooling process, not only for the AR loops, but also for the diffuse emission between the loops.
\citet{2013A&A...556A..79A} analyzed some coronal loops that were initially seen in the AIA 94\,\AA~channel, subsequently in the AIA 335\,\AA, and in one case in the AIA 211\,\AA~channel, and provided evidence of the cooling of impulsively heated loops.
By employing AIA multi-wavelength images, \citet{2015A&A...583A.109L} presented the cooling process of heated AR loops, and obtained the consistent time lags with the one-dimensional loop nanoflare model of \citet{2006ApJ...647.1452P}.

Magnetic flux rope is a coherent helical structure along the polarity inversion lines (PILs), with field lines wound around a central axis \citep{2016ApJ...818..148L, 2022ApJ...941L...1L}.
It is a multithermal feature, which could be a filament (prominence)\citep{2010SSRv..151..333M, 2013SoPh..282..147L, 2021ApJ...919L..21L}, warm channel \citep{2022ApJ...941L...1L, 2022ApJ...933...68S}, and hot channel \citep{2012ApJ...761...62C, 2016ApJ...829L..33L}.
Similar to the widely investigated hot channel, the warm channel here also shows a coherently elongated channel-like structure.
It, however, appears in the passbands that sample plasma at temperatures of around 1\,MK, rather than the high-temperature passbands in which the hot channel is observed \citep[see more details in][]{2022ApJ...941L...1L, 2022ApJ...933...68S}.
Flux ropes sometimes erupt \citep{2003A&A...397.1057Z, 2012ApJ...750...12S, 2013AJ....145..153Y, 2016NatPh..12..847L}.
Successful flux rope eruptions push the overlying loops outward, forming the bright compression fronts, that become the bright fronts of coronal mass ejections (CMEs) in the coronagraphs \citep{2012A&A...539A...7L, 2012ApJ...749...12Y, 2019ApJ...887..124S, 2022ApJ...933...68S}.
They also drive the nearby coronal loops, resulting in the kinetics, such as the expansion, contraction, and oscillation, of coronal loops \citep[e.g.,][]{2002SoPh..206...69S, 2011SoPh..270..191M, 2013MNRAS.431.1359Z, 2015ApJ...805....4Z, 2022ApJ...937L..21Z}.
Recently, by using AIA and the New Vacuum Solar Telescope \citep[NVST;][]{2014RAA....14..705L} images, \citet{2021ApJ...908..213L} reported that magnetic reconnection between loops is largely enhanced by a nearby filament eruption.

The legs, with opposite magnetic polarities, of the overlying loops that are stretched upward by the lower-lying erupting flux rope, converge \citep{1999Ap&SS.264..129S, 2000JGR...105.2375L}.
At the interface of the converging legs, the current sheets, connecting the upper erupting flux rope and the lower flare arcades, form \citep{2016ApJ...829L..33L, 2018ApJ...869..118D}.
In the current sheets, magnetic reconnection takes place, and converts  magnetic energy to other forms, e.g., kinetic and thermal \citep{2000mrmt.conf.....P, 2016NatPh..12..847L, 2021ApJ...908..213L}. 
The beamed nonthermal particles and/or thermal conduction front, produced during the reconnection, heat the chromospheric material quickly \citep{1999Ap&SS.264..129S, 2000JGR...105.2375L}.
This leads to the chromospheric evaporation and then the mass loading of the postflare loops \citep{2021RAA....21..255L, 2022FrASS...920116A}.
Bright flare ribbons and postflare loops then form \citep{2009ApJ...690..347L, 2012A&A...539A...7L}.
The hot postflare loops appear in the high-temperature passbands, e.g., soft X-rays and EUV 131 and 94\,\AA.
Subsequently, they cool down, and are seen sequentially in the low-temperature passbands, e.g., EUV 193, 171, and 304\,\AA~\citep{2021RAA....21...66L}.

The flares generally have two parallel ribbons, the footpoints of postflare loops \citep{2009ApJ...690..347L, 2021RAA....21...66L}.
Besides the standard two-ribbon flares, the circular-ribbon flares are frequently observed \citep[e.g.,][]{2009ApJ...700..559M, 2012ApJ...760..101W}. 
They comprise mainly a quasi-circular ribbon and an elongated inner ribbon enclosed, and sometimes several brightening patches, i.e., the remote ribbon, in the remote place outside the quasi-circular ribbon \citep{2013ApJ...778..139S, 2017ApJ...851...30X}.
The circular-ribbon flares are usually suggested to be the result of reconnection occurring at the coronal null points of the fan-spine topology, that could be naturally formed when a parasitic polarity (i.e., a minor polarity) is encompassed by a parent polarity (i.e., the opposite major polarity) on the photosphere \citep{1990ApJ...350..672L, 1998ApJ...502L.181A}. 
Heat flux and energized particles, produced during the reconnection, flow along the separatrices/quasi-separatrix layers (QSLs), and light up their footprints in the lower atmosphere.
The dome-shaped fan creates the quasi-circular ribbon, and spine maps to the inner and remote ribbons \citep{2013ApJ...778..139S, 2019ApJ...885L..11S}.
The postflare loops connect the ribbons that reside in opposite magnetic polarities.
Recently, several studies suggest that the reconnection at the coronal null points can be triggered by the flux rope eruption within the fan-spine topology \citep{2017ApJ...851...30X, 2020ApJ...900..158Y}.

In this paper, by employing observations from the SDO and the Solar Upper Transition Region Imager \citep[SUTRI;][]{2023RAA}, we investigate the heating of quiescent, i.e., non-flaring, coronal loops caused by nearby eruptions, and suggest that magnetic reconnection between the coronal loops triggered by the nearby eruptions heats the loops.
The observations, results, and summary and discussions are described in Sections\,\ref{sec:obs}, \ref{sec:res}, and \ref{sec:sum}, respectively.

\section{Observations}\label{sec:obs}

The SUTRI on board the Space Advanced Technology demonstration satellite (SATech-01), which was launched to a Sun-synchronous orbit at a height of $\sim$500\,km in July 2022 by the Chinese Academy of Sciences (CAS), aims to establish connections between structures in the lower solar atmosphere and corona, and advance our understanding of various types of solar activity, such as the flares, filament eruptions, and CMEs \citep{2023RAA}.
It takes full-disk solar atmospheric images, with a field of view (FOV) of $\sim$41.6\arcmin$\times$41.6\arcmin~and moderate spatial sampling of  $\sim$1.23\arcsec/pixel, at the Ne VII 465\,\AA~spectral line, formed at a temperature regime of $\sim$0.5\,MK in the solar atmosphere \citep{2017RAA....17..110T}, with a filter width of $\sim$30\,\AA.
Noting that SATech-01 is not a solar dedicated spacecraft, the solar observation time is $\sim$16\,hr each day because the earth eclipse time accounts for $\sim$1/3 of SATech-01's orbit period, and the normal time cadence of SUTRI images is $\sim$30\,s.

The SDO/AIA is a set of normal-incidence imaging telescopes, acquiring images of the solar atmosphere in 10 wavelength bands.
Different AIA channels sample plasma at different temperatures, e.g., 94\,\AA~peaks at $\sim$7.2\,MK (Fe XVIII), 335\,\AA~peaks at $\sim$2.5\,MK (Fe XVI), 211\,\AA~peaks at $\sim$1.9\,MK (Fe XIV), 193\,\AA~peaks at $\sim$1.5\,MK (Fe XII), 171\,\AA~peaks at $\sim$0.9\,MK (Fe IX), 131\,\AA~peaks at $\sim$0.6\,MK (Fe XIII) and $\sim$10\,MK (Fe XXI), and 304\,\AA~peaks at $\sim$0.05\,MK (He II).
Here, we should notice that each AIA channel has contributions from multiple lines at different temperatures, and is not a clean representation of the characteristic temperature \citep{2010A&A...521A..21O}.
The spatial sampling and time cadence of AIA EUV images are 0.6\,\arcsec/pixel and 12\,s, respectively.
The Helioseismic and Magnetic Imager \citep[HMI;][]{2012SoPh..275..229S} on board the SDO provides line-of-sight (LOS) magnetograms, with the time cadence and spatial sampling of 45\,s and 0.5\arcsec/pixel, respectively.

In this study, we use the SUTRI 465\,\AA~and AIA 131, 94, 335, 211, 193, 171, 304, and 1600\,\AA~images to investigate the evolution of eruptions and coronal loops.
Here, the AIA images are processed to 1.5 level uisng ``aia$\_$prep.pro".
To better show the evolution, the SUTRI and AIA EUV images are enhanced by using the Multiscale Gaussian Normalization (MGN) technique \citep{2014SoPh..289.2945M}.
We also employ the HMI LOS magnetograms to analyze the evolution of magnetic fields associated with the eruptions and coronal loops.

\section{Results}\label{sec:res}

On 2022 September 4, two ARs 13092 and 13093 were located near the southeastern solar limb; see the positive and negative magnetic fields P1 and N1, and P2 and N2 in Figure\,\ref{f:general_information}(a).
Furthermore, a negative magnetic field N3 between two ARs, and a positive field P3 located to the west of AR 13093 are observed.
A long U-shaped filament, F0, with a mean length of $\sim$650\,Mm, is detected in AIA and SUTRI EUV images; see Figures\,\ref{f:general_information}(b)-(d).
It is located upon the PILs between the positive and negative magnetic fields, P1, P2, and P3, and N1 and N3; see the green dashed line in Figure\,\ref{f:general_information}(a).
To the south of the long filament F0, a short sigmoidal filament, F2, is identified; see Figure\,\ref{f:general_information}(b1).
It is located upon the PIL between the positive and negative magnetic fields, P3 and N2; see the blue line in Figure\,\ref{f:general_information}(a).
Overlying the filament F0, quiescent coronal loops, L1, are identified in AIA 211\,\AA~images; see Figure\,\ref{f:general_information}(d1).
They connect the positive and negative magnetic fields P2 and N3.
Another coronal loops, L2, overlying the filament F2, are also detected in AIA 211\,\AA~images; see Figure\,\ref{f:general_information}(d1).
They connect the positive and negative magnetic fields P3 and N2.

Employing the potential field source surface (PFSS) model \citep{1969SoPh....6..442S} that provides an approximate description of the coronal magnetic field based on the observed photospheric magnetic fields (magnetograms), we derive the coronal magnetic field in a spherical shell spanning radial distances from 1.0 to 2.5 solar radii.
In Figure\,\ref{f:pfss_fields}, the red, green, and blue lines represent the PFSS coronal magnetic field lines connecting the positive and negative magnetic fields, P2 and N3, P3 and N3, and P2 and P3 and N2, respectively.
Here, the red and blue lines are separately consistent with the field lines of loops L1 and L2 in Figure\,\ref{f:general_information}(d1).

\subsection{Eruptions}\label{subsec:eruptions}

From $\sim$05:13 UT, a faint bright structure, F1, is detected in AIA low-temperature, e.g., 335, 211, and 193\,\AA, rather than high-temperature, e.g., 94 and 131\,\AA, images; see Figure\,\ref{f:filament_eruptions}(a1).
It erupted toward the northeast; see the online animated version of Figure\,\ref{f:filament_eruptions}.
The erupting structure shows a channel-like configuration with a mean temperature of $\sim$2\,MK, consistent with the warm channel reported in \citet{2022ApJ...941L...1L}.
Along the AB direction in the blue rectangle in Figure\,\ref{f:filament_eruptions}(a1), a time slice of AIA 211\,\AA~images is made, and displayed in Figure\,\ref{f:time_slices1}(a).
The warm channel F1 began to rise slowly with a mean speed of $\sim$2\,km\,s$^{-1}$, and then quickly erupted with a mean acceleration and speed of $\sim$6\,m\,s$^{-2}$ and $\sim$26\,km\,s$^{-1}$, respectively; see the blue dotted line in Figure\,\ref{f:time_slices1}(a).
Two bright ribbons appeared on both sides of the erupting warm channel F1, marked by two blue solid arrows in Figures\,\ref{f:filament_eruptions}(b1)-(b3) and (d1).
Postflare loops, connecting the ribbons, formed in AIA EUV images, marked by the purple solid arrows in Figures\,\ref{f:heating_evolution}(a)-(d).

From $\sim$05:30 UT, material of the warm channel (also marked by F1), seen in AIA and SUTRI EUV images, began to move toward the northwest; see Figure\,\ref{f:filament_eruptions}(b1).
It passed under the filament F0; see Figures\,\ref{f:filament_eruptions}(b2) and (d1), and then turned to the west; see Figure\,\ref{f:filament_eruptions}(b3).
Subsequently, it went over the filament F0, divided into two branches, denoted by the green solid arrows in Figure\,\ref{f:filament_eruptions}(b3), and fell northward and southward to the solar surface; see the online animated version of Figure\,\ref{f:filament_eruptions}.
Along the CD direction in the blue rectangle in Figure\,\ref{f:filament_eruptions}(b2), a time slice of AIA 304\,\AA~images is made, and displayed in Figure\,\ref{f:time_slices1}(b).
The material of warm channel F1 moves with a mean acceleration of $\sim$6\,m\,s$^{-2}$ and speed of $\sim$38\,km\,s$^{-1}$; see the green dotted line in Figure\,\ref{f:time_slices1}(b).
Combining the AIA EUV images with the HMI LOS magnetograms, we notice that the warm channel F1 is an elongated structure, located upon the PILs between the positive and negative magnetic fields P2 and N2, and P3 and N3.
Furthermore, only the short southeastern part, i.e., the warm channel in Figure\,\ref{f:filament_eruptions}(a1), rather than the long northwestern part, tracked by the moving material in Figures\,\ref{f:filament_eruptions}(b1)-(b3), of the elongated structure erupted.

From $\sim$07:17 UT, the filament F2 erupted toward the southeast; see Figures\,\ref{f:filament_eruptions}(b4)-(b6) and (d2).
This filament eruption may be driven by the weakening of overlying loops caused by the previous eruption of warm channel F1.
Along the EF direction in the blue rectangle in Figure\,\ref{f:filament_eruptions}(b5), a time slice of AIA 304\,\AA~images is made, and shown in Figure\,\ref{f:time_slices1}(c).
The green dotted line in Figure\,\ref{f:time_slices1}(c) outlines the eruption of filament F2, with a mean acceleration and speed of $\sim$15\,m\,s$^{-2}$ and $\sim$62\,km\,s$^{-1}$.
Two flare ribbons and postflare loops formed underneath the filament eruption; see Figures\,\ref{f:filament_eruptions}(b4)-(b6), (c), and (d2), and the online animated version of Figure\,\ref{f:filament_eruptions}.

The propagation of material of warm channel F1 under the filament F0 pushed the material of filament F0 to the northwest, with a mean speed of $\sim$9\,km\,s$^{-1}$; see the cyan dotted line in Figure\,\ref{f:time_slices1}(b), that moved back due to the magnetic tension force of filament F0, showing a kink oscillation of filament F0.
The eruption of filament F2 then enhanced the oscillation.
Along the GH direction in the blue rectangle in Figure\,\ref{f:filament_eruptions}(c), a time slice of AIA 171\,\AA~images is made, and displayed in Figure\,\ref{f:time_slices1}(d).
The left green solid arrow denotes the oscillation of filament F0, with an amplitude of $\sim$16\,Mm, caused by the propagation of material of warm channel F1, denoted by the green dotted line in Figure\,\ref{f:time_slices1}(d).
After the eruption of filament F2, marked by the blue vertical dashed line in Figure\,\ref{f:time_slices1}(d), the oscillation was enhanced with a larger amplitude of $\sim$40\,Mm; see the second green solid arrow on the left in Figure\,\ref{f:time_slices1}(d).
Subsequently, the oscillation faded away, and finally stopped, showing a damped oscillation, with a mean period of $\sim$1.4\,hr; see the green solid arrows in Figure\,\ref{f:time_slices1}(d).
Moreover, oscillations of multiple threads of the filament F0 are identified in Figure\,\ref{f:time_slices1}(d).

\subsection{Heating of coronal loops}\label{subsec:heating_loops}

After the eruption of warm channel F1, to the northeast of the associated flare, coronal loops L1, with a length of $\sim$140\,Mm, appeared in the AIA high-temperature channels, e.g., 94 and 131\,\AA; see Figures\,\ref{f:loop_heating}(a) and (b).
They are not observed at the same time, and will be seen during the subsequent cooling process; see Section\,\ref{subsec:cooling_loops}, in the AIA low-temperature channels, e.g., 335, 211, 193, 171, and 304\,\AA, and the SUTRI 465\,\AA~channel, which samples plasma with the characteristic temperature of $\sim$0.5\,MK, similar to the lower characteristic temperature ($\sim$0.6\,MK) of the AIA 131\,\AA~channel.
The loops L1 are hence heated to a higher temperature associated with the nearby warm channel eruption.
Comparing Figures\,\ref{f:loop_heating}(a) and (b), we notice that more loops are observed in the AIA 94\,\AA~channel than in the 131\,\AA~channel.
This suggests that more loops are heated to $\sim$7.2\,MK, the characteristic temperature of the AIA 94\,\AA~channel, and less loops are heated to $\sim$10\,MK, the higher characteristic temperature of the AIA 131\,\AA~channel.

Using six AIA EUV channels, including 94, 335, 211, 193, 171, and 131\,\AA, we analyze the temperature and EM of the loops L1.
Here, we employ the differential EM (DEM) analysis using ``xrt$\_$dem$\_$iterative2.pro" \citep{2004IAUS..223..321W, 2012ApJ...761...62C}.
The loop region, enclosed by the red rectangles in Figures\,\ref{f:loop_heating}(a) and (b), is chosen to compute the DEM.
The region, enclosed by the pink rectangles in Figures\,\ref{f:loop_heating}(a) and (b), out of the loops L1 is chosen for the background emission that is subtracted from the loop region.
In each region, the digital number counts in each of the six AIA EUV channels are temporally normalized by the exposure time and spatially averaged over all pixels.
The DEM curve of the loop region is displayed in Figure\,\ref{f:dems}(a).
The average DEM-weighted temperature and EM are 8.2\,MK and 4.9$\times$10$^{27}$\,cm$^{-5}$, respectively.
This further supports that the loops L1 are heated.

Employing the EM, the electron number density ($n_{e}$) of the loops L1 is estimated using $n_{e}$=$\sqrt{\frac{EM}{D}}$, where $D$ is the LOS depth of the loops.
Assuming that the depth $D$ equals the width ($W$) of the loops, then the electron number density is $n_{e}$=$\sqrt{\frac{EM}{W}}$.
We measure the width of loops L1 at the place and time, where and when we calculate the DEM.
First we get the intensity profile in the AIA 131\,\AA~channel perpendicular to the loops L1.
We use the mean intensity surrounding the loops as the background emission, and subtract it from the intensity profile.
Employing a single Gaussian, we fit the residual intensity profile, and achieve the FWHM of the single Gaussian fit as the width.
The width of the loops L1 is measured to be $\sim$4.4\,Mm.
Using EM=4.9$\times$10$^{27}$\,cm$^{-5}$ and $W$=4.4\,Mm, we obtain the electron number density of the loops L1 to be 3.3$\times$10$^{9}$\,cm$^{-3}$.

In the white rectangles in Figures\,\ref{f:loop_heating}(a) and (b), light curves of the AIA 94 and 131\,\AA~channels are calculated, and shown in Figures\,\ref{f:lightcurves1}(a) and (b), respectively.
The AIA 94\,\AA~light curve started to increase slowly from $\sim$05:37 UT; see the blue vertical dotted line in Figure\,\ref{f:lightcurves1}(b).
Subsequently, the AIA 131\,\AA~light curve began to increase from $\sim$05:50 UT, since which the AIA 94\,\AA~light curve increase quickly; see the green vertical dotted lines in Figures\,\ref{f:lightcurves1}(a) and (b).
This may imply that the loops L1 are heated first to $\sim$7.2\,MK, and part of them are then heated to $\sim$10\,MK.
There is another possibility that the loops L1 consist of many strands, and different strands get heated to different temperatures.
Moreover, as the AIA 131\,\AA~channel clearly shows a strong cool background that is not present in the AIA 94\,\AA~channel, and the background intensity level is higher, it is also possible that the increase of the AIA 94\,\AA~light curve is detected earlier because of a higher signal-to-noise.
The AIA 131\,\AA~light curve reaches the peak at $\sim$06:29 UT, $\sim$14 minutes earlier than the peak ($\sim$06:43 UT) of the AIA 94\,\AA~light curve; see the blue vertical dashed lines in Figures\,\ref{f:lightcurves1}(a) and (b).
Moreover, the duration of the increase of the AIA 94\,\AA~light curve is longer than that of the AIA 131\,\AA~light curve, supporting that more loops (strands) are heated to $\sim$7.2\,MK, and sustain for longer time.
Before the eruption of filament F2, the AIA 131\,\AA~loops disappeared, while the AIA 94\,\AA~loops remained; see the online animated version of Figure\,\ref{f:loop_heating}.

After the eruption of filament F2, to the east of the associated flare, the loops L1 re-appeared in AIA 131\,\AA~images, and became more in AIA 94\,\AA~images.
Similar to the previous heating process of loops L1, no loop is observed at the same time in AIA low-temperature, e.g., 193 and 171\,\AA, and SUTRI 465\,\AA~images.
This indicates that the loops L1 are heated again associated with the nearby filament eruption.
We also analyze the temperature and EM of loops L1 during the second heating process.
The red and pink rectangles in Figures\,\ref{f:loop_heating}(c) and (d) separately enclose the loop region where we compute the DEM and the region out of the loops where we calculate the background emission.
The DEM curve of the loop region is distributed in Figure\,\ref{f:dems}(b).
The average DEM-weighted temperature and EM are 8.5 MK and 1.4$\times$10$^{28}$\,cm$^{-5}$, respectively.
The width of loops L1 at the place and time, where and when we compute the DEM, is measured to be $\sim$6\,Mm.
Employing EM=1.4$\times$10$^{28}$\,cm$^{-5}$ and $W$=6\,Mm, we achieve the electron number density of loops L1 to be 4.8$\times$10$^{9}$\,cm$^{-3}$.
The loops L1 are thus hotter and denser during the second heating process than those during the first heating process.

The second heating process of loops L1 is also shown in the light curves of the AIA 131 and 94\,\AA~channels as displayed in Figures\,\ref{f:lightcurves1}(a) and (b).
The AIA 131 and 94\,\AA~light curves both increase from $\sim$07:31 UT, and peak at $\sim$07:45\,UT; see the red vertical dotted and dashed lines in Figures\,\ref{f:lightcurves1}(a) and (b).
Comparing with those during the first heating process, they reach the higher peaks in shorter times, i.e., 14\,minutes.
The AIA 94\,\AA~loops sustain for a longer time than the AIA 131\,\AA~loops, and both of them then disappear.

The detailed evolution of the heated loops in AIA 94\,\AA~images is investigated; see Figures\,\ref{f:heating_evolution}(a)-(d).
The loops L1 connect the positive and negative magnetic fields P2 and N3; enclosed by the red and blue rectangles in Figures\,\ref{f:heating_evolution}(c) and (h).
To the west of loops L1, another loops, L2, connecting the positive and negative magnetic fields, P3 and N2, are detected; see Figure\,\ref{f:pfss_fields}.
Moreover, postflare loops associated with the eruptions of warm channel F1 and filament F2, separately denoted by the purple and pink solid arrows in Figures\,\ref{f:heating_evolution}(a)-(d), are identified.

During the first heating process of loops, the loops L1 moved toward the loops L2, constituting an X-type structure a little above the southern endpoint of loops L1; see Figure\,\ref{f:heating_evolution}(a).
They then disappeared at the interface, with the formation of loops L3, connecting the positive and negative magnetic fields P3 and N3; see the blue circle and rectangle in Figures\,\ref{f:heating_evolution}(a)-(d).
This suggests that magnetic reconnection occurs between the loops L1 and L2.
Along the IJ direction in the pink rectangle in Figure\,\ref{f:heating_evolution}(a), a time slice of AIA 94\,\AA~images is made, and displayed in Figure\,\ref{f:time_slices2}(a).
The motion of loops L1 toward the loops L2 is clearly detected, with a mean speed of $\sim$14\,km\,s$^{-1}$; see the red dotted line in Figure\,\ref{f:time_slices2}(a).
At the meantime, the loops L1 propagated from west to east at the northern endpoints; see Figures\,\ref{f:heating_evolution}(a)-(d), due to the progressive heating of loops L1.
Subsequently, similar evolution of loops L1 and L2 took place at a higher position of loops L1; see Figure\,\ref{f:heating_evolution}(b).
Along the KL direction in the pink rectangle in Figure\,\ref{f:heating_evolution}(b), we make a time slice of AIA 94\,\AA~images, and show it in Figure\,\ref{f:time_slices2}(b).
Motion of loops L1 toward the loops L2, similar to Figure\,\ref{f:time_slices2}(a), is evidently identified, with a smaller mean speed of $\sim$8\,km\,s$^{-1}$; see the red dotted line in Figure\,\ref{f:time_slices2}(b).
Besides the newly reconnected loops L3, the newly formed loops L4, connecting the positive and negative magnetic fields P2 and N2, enclosed by two red circles in Figures\,\ref{f:heating_evolution}(c) and (h), are detected.

During the second heating process of loops, the loops L2 overlying the filament F2 are observed; see Figure\,\ref{f:heating_evolution}(d).
Along with the eruption of filament F2, they are pushed to the southeast, and meet the loops L1, with the formation of loops L3; see the online animated version of Figure\,\ref{f:loop_heating}.
Along the MN direction in the pink rectangle in Figure\,\ref{f:heating_evolution}(d), a time slice of AIA 94\,\AA~images is made, and shown in Figure\,\ref{f:time_slices2}(c).
The pink dotted line in Figure\,\ref{f:time_slices2}(c) denotes the eruption of filament F2, with a mean speed of $\sim$53\,km\,s$^{-1}$.
The red dotted line in Figure\,\ref{f:time_slices2}(c) outlines the propagation of loops L2 toward the loops L1, with a mean acceleration and speed of $\sim$31\,m\,s$^{-2}$ and $\sim$50\,km\,s$^{-1}$.

In AIA and SUTRI images, brightening appeared at the endpoints of newly reconnected loops L3, enclosed by the blue rectangle and circle in Figure\,\ref{f:heating_evolution}.
The brightening at the northern endpoints of loops L1 (L3) moved from west to east during the heating process of loops L1; see Figures\,\ref{f:heating_evolution}(e) and (f), and the online animated version of Figure\,\ref{f:filament_eruptions}, similar to the evolution of loops L1.
In the blue rectangles in Figures\,\ref{f:heating_evolution}(e)-(g), light curves of the AIA 304 and 1600\,\AA~channels are calculated, and displayed in Figures\,\ref{f:lightcurves1}(c) and (d), respectively.
The AIA 304 and 1600\,\AA~light curves have the similar evolution to the AIA 94\,\AA~light curve; see Figures\,\ref{f:lightcurves1}(b)-(d).

\subsection{Cooling of coronal loops}\label{subsec:cooling_loops}

After the appearance of the heated loops L1, (at least) three sets of loops are seen at the same place sequentially in the AIA low-temperature channels; see the online animated version of Figure\,\ref{f:loop_cooling}.
We choose one of them, and calculate the temperature and EM of the loops L1 in the loop region enclosed by the red rectangle in Figure\,\ref{f:loop_cooling}(e).
In the region out of the loops L1, enclosed by the pink rectangle in Figure\,\ref{f:loop_cooling}(e), the background emission is measured.
The DEM curve of the loop region is displayed in Figure\,\ref{f:dems}(c).
Compared with those in Figures\,\ref{f:dems}(a) and (b), the emission of loops L1 here is mainly from plasma with lower temperatures.
The mean DEM-weighted EM and temperature are separately 2.2$\times$10$^{27}$\,cm$^{-5}$ and 1.6 MK. 
The width of loops L1 in the AIA 171\,\AA~channel at the place and time, where and when the DEM curve is computed, is measured to be 6.4\,Mm.
Employing EM=2.2$\times$10$^{27}$\,cm$^{-5}$ and $W$=6.4\,Mm, the electron number density of loops L1 is estimated to be 1.9$\times$10$^{9}$\,cm$^{-3}$.
Compared with the heated loops, the loops L1 are cooler and rarer.
Moreover, similar results are obtained for other two sets of cooling loops.

In the black rectangles in Figure\,\ref{f:loop_cooling}, the light curves of the loops of the AIA and SUTRI EUV channels are calculated, and shown in Figure\,\ref{f:lightcurves2}.
The AIA 131 and 94\,\AA~light curves in Figures\,\ref{f:lightcurves2}(a) and (b) are slightly different from those in Figures\,\ref{f:lightcurves1}(a) and (b) during the heating processes, as the chosen loop regions are different.
When the light curves of the AIA 131 and 94\,\AA~channels increase from $\sim$05:46 UT, those of the AIA 335 and 211\,\AA~channels decrease clearly; see the blue vertical dotted lines in Figures\,\ref{f:lightcurves2}(a)-(d).
The AIA 335\,\AA~light curve then increases, and reaches the peak at $\sim$07:24 UT, after the first peaks of the AIA 131 ($\sim$06:33 UT) and 94\,\AA~($\sim$06:52 UT) light curves; see the blue vertical dashed lines in Figures\,\ref{f:lightcurves2}(a)-(c).
This indicates that the heated loops L1 cooled down to $\sim$2.5\,MK, the characteristic temperature of the AIA 335\,\AA~channel.
The light curves of the AIA 211 and 193\,\AA~channels increase slightly before the second peaks of the AIA 131 and 94\,\AA~light curves; see Figures\,\ref{f:lightcurves2}(d) and (e), showing that less heated loops cooled down further to $\sim$1.5\,MK, the characteristic temperature of the AIA 193\,\AA~channel; see also  the online animated version of Figure\,\ref{f:loop_cooling}.

After the second peaks ($\sim$07:45 UT) of the AIA 131 and 94\,\AA~light curves, the light curves of the AIA 335, 211, 193, and 171\,\AA~channels increase, and reach the peaks at $\sim$09:47, $\sim$10:23, $\sim$10:32, and $\sim$10:46 UT, respectively; see the red vertical dashed lines in Figures\,\ref{f:lightcurves2}(a)-(f).
The third peaks of the AIA 94 ($\sim$10:46 UT) and 131\,\AA~($\sim$10:48 UT) light curves, denoted by the purple vertical dashed lines in Figures\,\ref{f:lightcurves2}(a) and (b), represent the emission from plasma with the lower characteristic temperatures of the AIA 94 and 131\,\AA~channels.
They also show the cooling process of the heated loops L1.
Moreover, we overlay the light curve of the SUTRI 465\,\AA~channel, with the characteristic temperature of $\sim$0.5\,MK, in Figure\,\ref{f:lightcurves2}(g) on Figure\,\ref{f:lightcurves2}(a) as red pluses.
No SUTRI 465\,\AA~loop is observed during the heating processes of loops L1.
After the appearance of loops L1 in SUTRI images, the SUTRI 465\,\AA~light curve coincides well with the AIA 131\,\AA~light curve.
This supports that the characteristic temperature of the SUTRI 465\,\AA~channel is similar to the lower one of the AIA 131\,\AA~channel.
Furthermore, the evolution of the other two sets of cooling loops is similar to that of the chosen loops, but with time shifts.

\section{Summary and discussion}\label{sec:sum}

Employing SDO/AIA and SUTRI images and SDO/HMI LOS magnetograms, we analyze the eruptions of warm channel F1 and filament F2, and the heating of quiescent loops L1 overlying the nearby filament F0.
On 2022 September 4, two neighboring ARs 13092 and 13093 are observed near the southeastern solar limb.
In these two ARs, the long U-shaped filament F0 is located upon the PILs among the positive and negative magnetic fields, P1, P2, and P3, and N1 and N3.
In AR 13093, the short sigmoidal filament F2 is located upon the PILs between the positive and negative magnetic fields P3 and N2.
The loops L1, overlying the filament F0, connect the positive and negative magnetic fields P2 and N3.
The loops L2, overlying the filament F2, connect the negative and positive magnetic fields, N2 and P3.
The southeastern part of the warm channel F1 in AR 13093 erupted toward the northeast in AIA low-temperature, e.g., 335, 211, and 193\,\AA, images.
Its material then moved to the northwest under the filament F0, turned to the west above the filament F0, divided into two branches, and fell to the solar surface northward and southward.
Subsequently, the filament F2 erupted to the southeast.
The two eruptions result in the decaying kink oscillation of nearby filament F0, with a mean oscillating period of $\sim$1.4\,hr.
During each eruption, the nearby quiescent loops L1 appeared in AIA high-temperature, e.g., 94 and 131\,\AA, images, with a mean DEM-weighted temperature of more than 8\,MK, indicating the heating of loops.
During the heating process, the loops L1 moved toward the loops L2, forming an X-type structure.
They disappeared at the interface, with the formation of two new loops L3 and L4, separately connecting the positive and negative magnetic fields, P3 and N3, and P2 and N2.
This suggests that magnetic reconnection took place between the loops L1 and L2.
The heated loops L1 then cooled down, and appeared sequentially in AIA and SUTRI lower-temperature, e.g., AIA 335, 211, 193, 171, 131\,\AA~and SUTRI 465\,\AA, images.

According to the AIA and SUTRI images, the HMI LOS magnetograms, and the PFSS coronal magnetic field, schematic diagrams are provided in Figure\,\ref{f:cartoon} to describe the heating of quiescent loops L1 caused by the nearby eruptions of warm channel F1 and filament F2.
The eruption of warm channel F1 toward the northeast; see the yellow lines and arrow in Figure\,\ref{f:cartoon}(a), pushes the nearby loops L1; see the red lines in Figure\,\ref{f:cartoon}(a), sideward to the west.
The loops L1 hence meet their neighboring loops L2; see the blue lines in Figure\,\ref{f:cartoon}(a), and reconnect with them; see the purple star in Figure\,\ref{f:cartoon}(b).
The newly reconnected loops L3 and L4; see the red blue lines in Figure\,\ref{f:cartoon}(b), then form.
The subsequent eruption of filament F2 toward the southeast; see the cyan line and arrow in Figure\,\ref{f:cartoon}(a), pushes the overlying loops L2 to the east, leading to another magnetic reconnection between the loops L1 and L2.
The thermal conduction front and/or beamed nonthermal particles, produced during both reconnection processes, propagate to the chromosphere along the newly reconnected loops L3 and L4.
They quickly heat the chromospheric material surrounding the loop endpoints, resulting in chromospheric evaporation and greater mass loading of the loops.
The loops L1, and also L3, then appear in AIA high-temperature images.
After the reconnection stops, the heated loops cool down, and appear sequentially in the AIA and SUTRI lower-temperature channels.

Heating of quiescent coronal loops caused by nearby eruptions is reported.
Quiescent, i.e., non-flaring, loops are usually thought to be heated by nanoflares \citep[e.g.,][]{1988ApJ...330..474P, 2006ApJ...647.1452P, 2010ApJ...723..713T, 2015A&A...583A.109L}.
In this study, quiescent loops appeared during the nearby eruptions of warm channel F1 and filament F2 in the AIA high-temperature rather than low-temperature channels; see Section\,\ref{subsec:heating_loops}.
This indicates the heating of loops, which is also supported by the DEM curves of loops; see Figure\,\ref{f:dems}.
More AIA 94\,\AA~loops than 131\,\AA~loops are detected; see Section\,\ref{subsec:heating_loops}.
This shows that more loops (strands) are heated to $\sim$7.2\,MK rather than $\sim$10\,MK, the characteristic temperatures of the AIA 94 and 131\,\AA~channels.
During the first heating process, the evident decrease of the AIA 335 and 211\,\AA~light curves suggests that the loops may be heated from $\sim$2 MK to more than 7\,MK, consistent with those in \citet{2015A&A...583A.109L}.
Moreover, signature of magnetic reconnection between loops L1 and L2 is identified during the heating process; see Section\,\ref{subsec:heating_loops}.
These results indicate that the loops may be heated by reconnection between loops, which are pushed by the underlying filament (flux rope) eruptions \citep{2002SoPh..206...69S, 2021ApJ...908..213L, 2022ApJ...937L..21Z}; see Figure\,\ref{f:cartoon}.
The simultaneous brightening in AIA 304 and 1600\,\AA~images at the loop footpoints provides the evidence of chromospheric evaporation induced by the reconnection between loops \citep{2021RAA....21...66L}.
The brightening propagation of the loops L1 in AIA 94 and 131\,\AA~images, and also their northern footpoints in AIA 304 and 1600\,\AA~images from west to east shows the successive reconnection between loops L1 and L2.

\begin{figure}[ht!]
\centering
\plotone{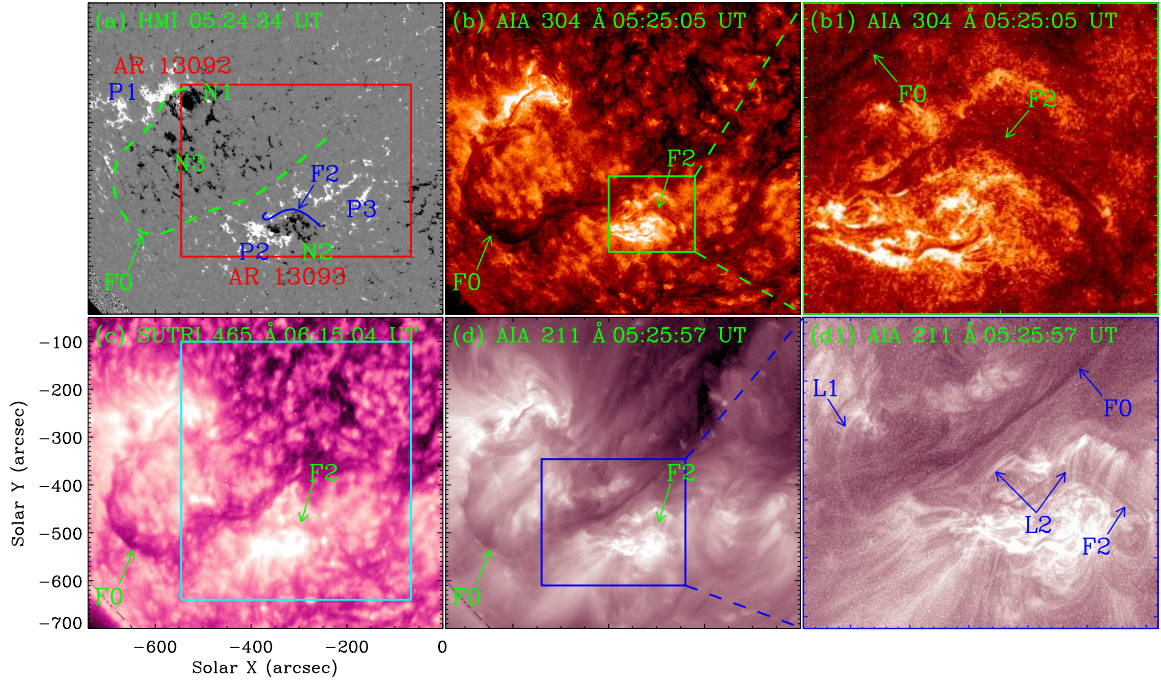}
\caption{General information of the heating of quiescent coronal loops caused by nearby eruptions.
(a) a SDO/HMI LOS magnetogram, and (b) and (b1) SDO/AIA 304\,\AA, (c) SUTRI 465\,\AA, and (d) and (d1) AIA 211\,\AA~images.
The AIA images in (b1) and (d1) have been enhanced by using the MGN technique. 
The green dashed and blue solid lines in (a) represent the filaments F0 and F2 in (b)-(d).
P1, P2, and P3, and N1, N2, and N3 in (a) mark the positive and negative magnetic fields.
The red and cyan rectangles in (a) and (c) separately show the FOVs of Figures\,\ref{f:pfss_fields} and \ref{f:cartoon} and Figures\,\ref{f:filament_eruptions}(a)-(c) and \ref{f:loop_heating}.
The green and blue rectangles in (b) and (d) denote the FOVs of (b1) and (d1), respectively.
L1 and L2 in (d1) illustrate the loops.
See Section\,\ref{sec:res} for details.
\label{f:general_information}}
\end{figure}

\begin{figure}[ht!]
\centering
\plotone{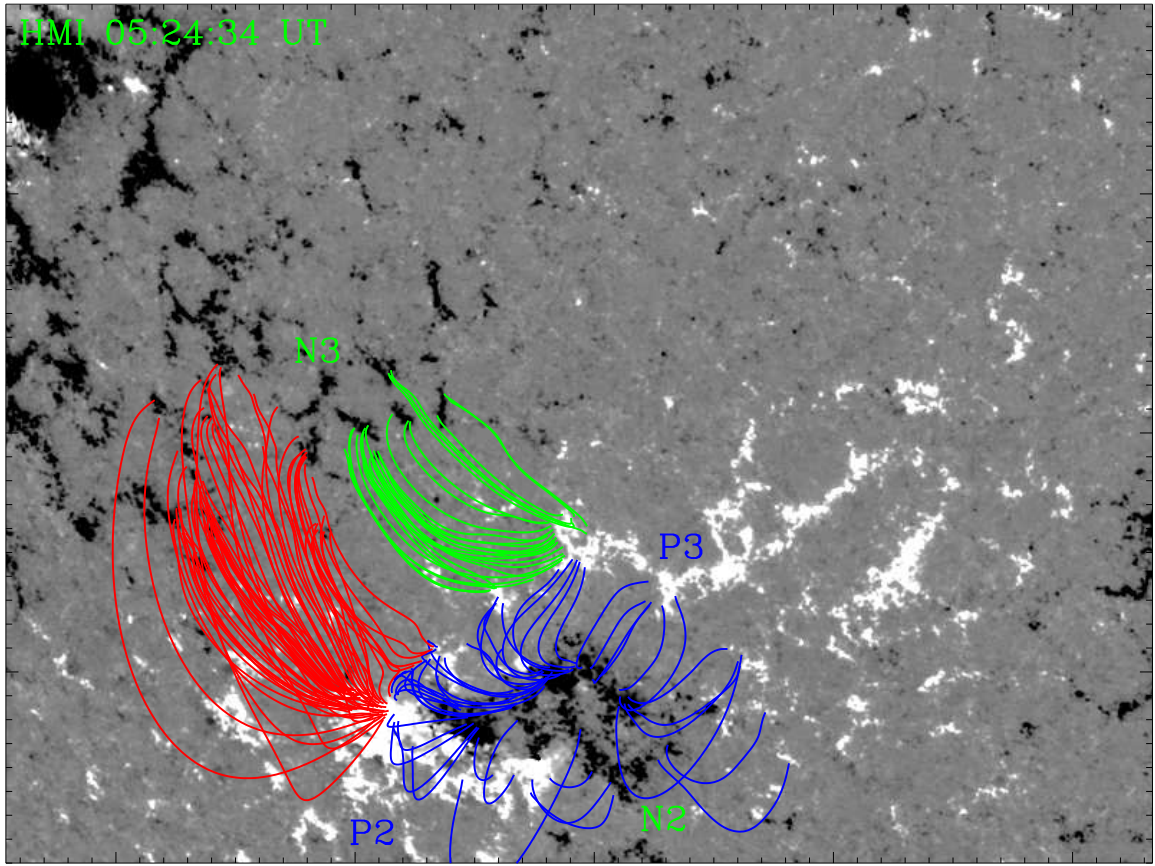}
\caption{Coronal magnetic field lines derived from the PFSS model. 
Similar to Figure\,\ref{f:general_information}(a), P2 and P3, and N2 and N3 mark the positive and negative magnetic fields, respectively.
The red, green, and blue lines represent the PFSS coronal magnetic field lines.
The FOV is denoted by the red rectangle in Figure\,\ref{f:general_information}(a).
See Section\,\ref{sec:res} for details.
\label{f:pfss_fields}}
\end{figure}

\begin{figure}[ht!]
\plotone{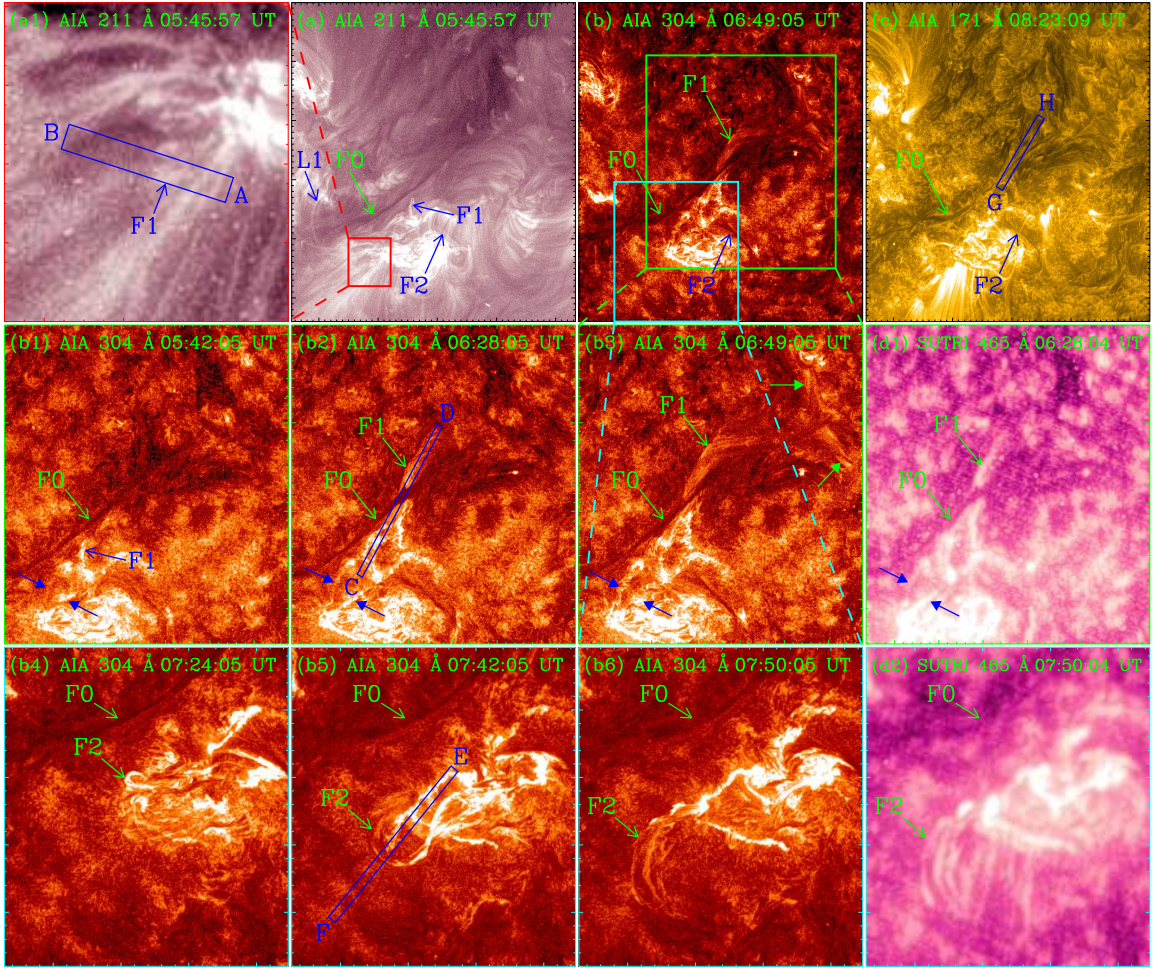}
\centering
\caption{Eruptions of the warm channel F1 and filament F2.
(a1) and (a) AIA 211\,\AA, (b) and (b1)-(b6) 304\,\AA, (c) 171\,\AA,  and (d1) and (d2) SUTRI 465\,\AA~images, enhanced by the MGN technique.
The red rectangle in (a) marks the FOV of (a1), and the green and cyan rectangles in (b) separately show the FOVs of (b1)-(b3) and (d1), and (b4)-(b6) and (d2).
The blue rectangles AB, CD, EF, and GH in (a1), (b2), (b5), and (c) mark the positions for the time slices in Figures\,\ref{f:time_slices1}(a)-(d), respectively.
The F1 in (a)-(b), (b1)-(b3), and (d1) mark the material, moving toward the northwest, of the erupting warm channel F1 in (a1).
The blue solid arrows in (b1)-(b3) and (d1) denote the brightenings.
The green solid arrows in (b3) mark two branches of warm channel F1.
An animation of the unannotated AIA images (panels (a)-(c)) is available.
It covers $\sim$9\,hr starting at 05:00 UT, with a time cadence of 1 minute.
The FOVs of (a)-(c) are indicated by the cyan rectangle in Figure\,\ref{f:general_information}(c).
See Section\,\ref{sec:res} for details.
(An animation of this figure is available.) 
\label{f:filament_eruptions}}
\end{figure}

\begin{figure}[ht!]
\plotone{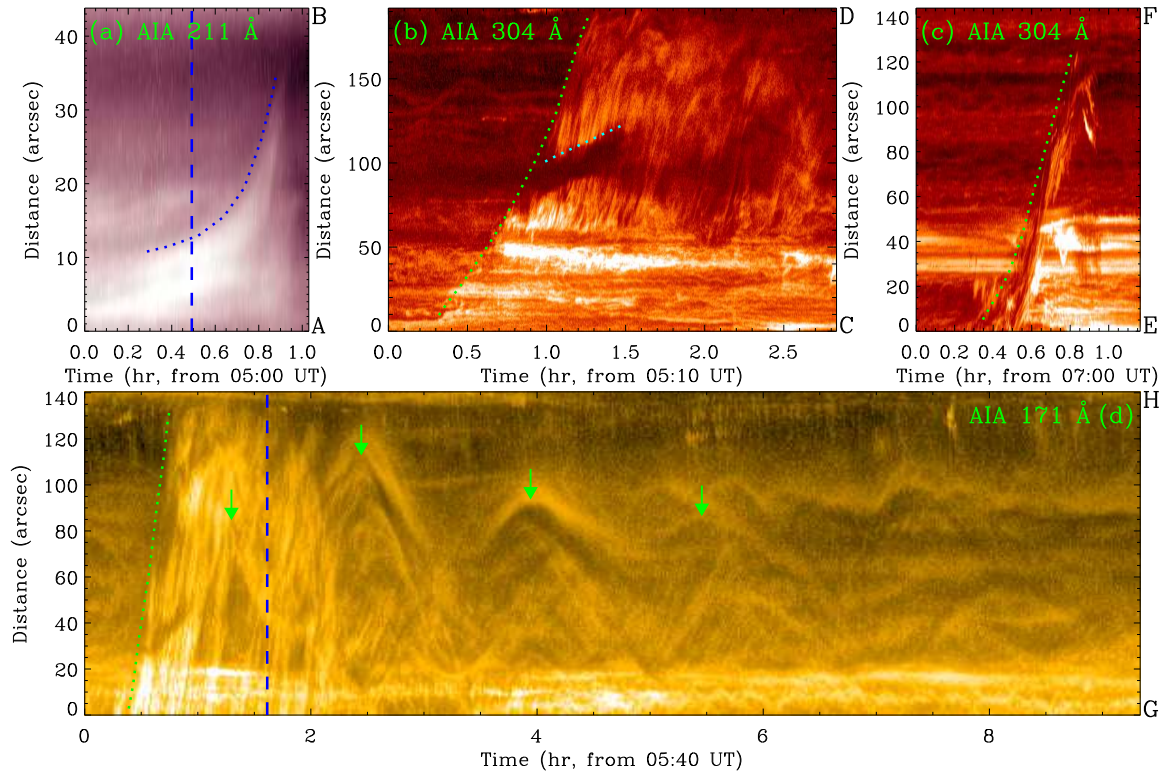}
\centering
\caption{Temporal evolution of the eruptions.
(a)-(d) Time slices of AIA 211\,\AA, 304\,\AA, and 171\,\AA~images along the directions AB, CD, EF, and GH in the blue rectangles in Figures\,\ref{f:filament_eruptions}(a1), (b2), (b5), and (c), respectively.
The blue and green dotted lines in (a)-(d) outline the eruptions.
The blue vertical dashed lines in (a) and (d) separately denote the start times of the evolution in (b) and (c).
The cyan dotted line in (b) and the green solid arrows in (d) mark the motion of filament F0.
See Section\,\ref{sec:res} for details.
\label{f:time_slices1}}
\end{figure}

\begin{figure}[ht!]
\centering
\includegraphics[width=0.8\textwidth]{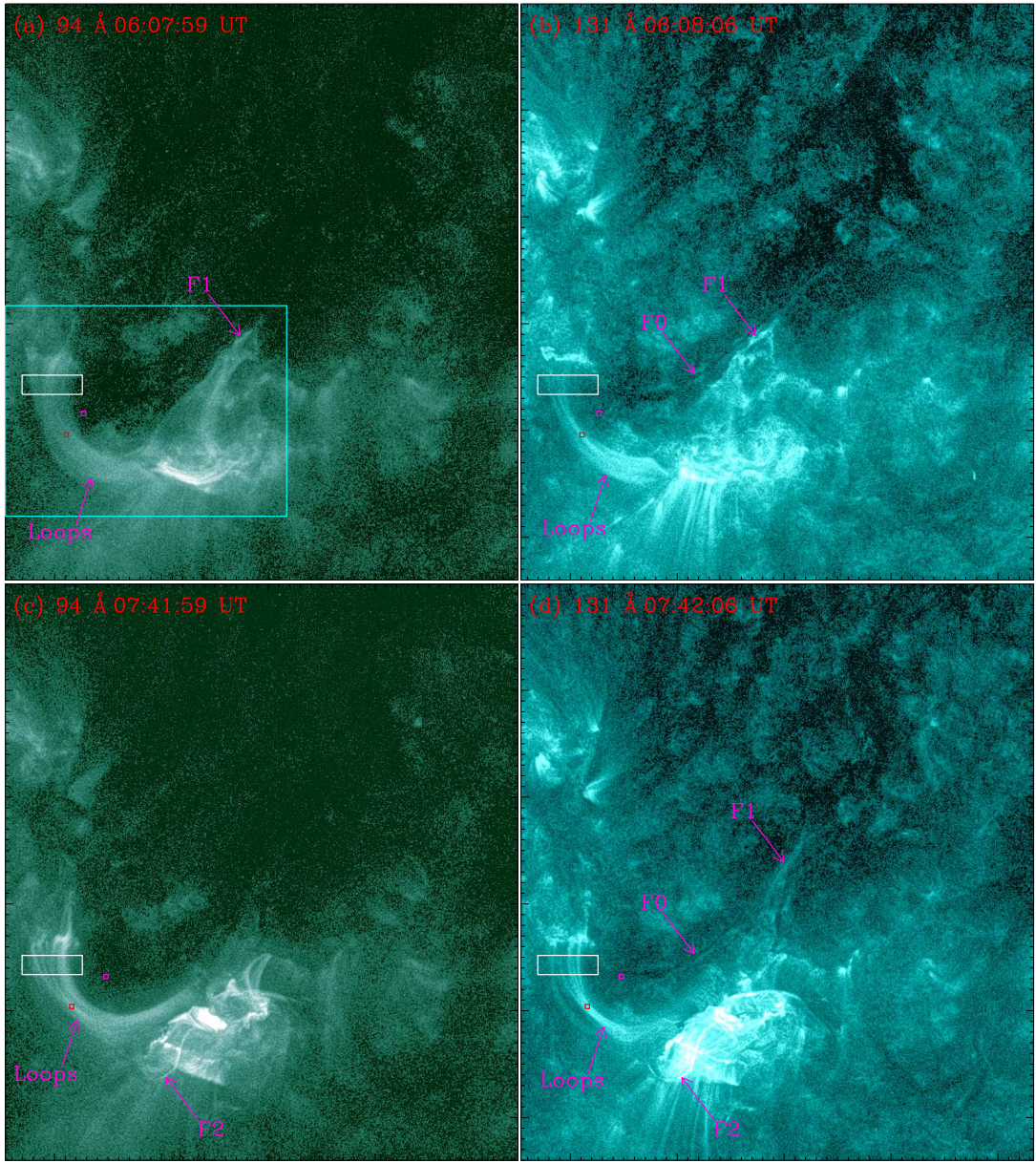}
\caption{Heating of the quiescent loops caused by nearby eruptions.
(a) and (c) AIA 94\,\AA~and (b) and (d) 131\,\AA~images, enhanced by the MGN technique.
The cyan rectangle in (a) shows the FOVs of Figures\,\ref{f:heating_evolution} and \,\ref{f:loop_cooling}.
The white rectangles in (a)-(d) mark the regions for the light curves in Figures\,\ref{f:lightcurves1}(a)-(b).
The red and pink rectangles in (a)-(d) enclose the regions for the DEM curves in Figures\,\ref{f:dems}(a)-(b), and the locations where the background emission is computed, respectively.
An animation of the unannotated AIA images (panels (a) and (b)) is available.
It covers $\sim$9\,hr starting at 05:00 UT, with a time cadence of 1 minute.
The FOV is indicated by the cyan rectangle in Figure\,\ref{f:general_information}(c).
See Section\,\ref{sec:res} for details.
(An animation of this figure is available.) 
\label{f:loop_heating}}
\end{figure}

\begin{figure}[ht!]
\centering
\includegraphics{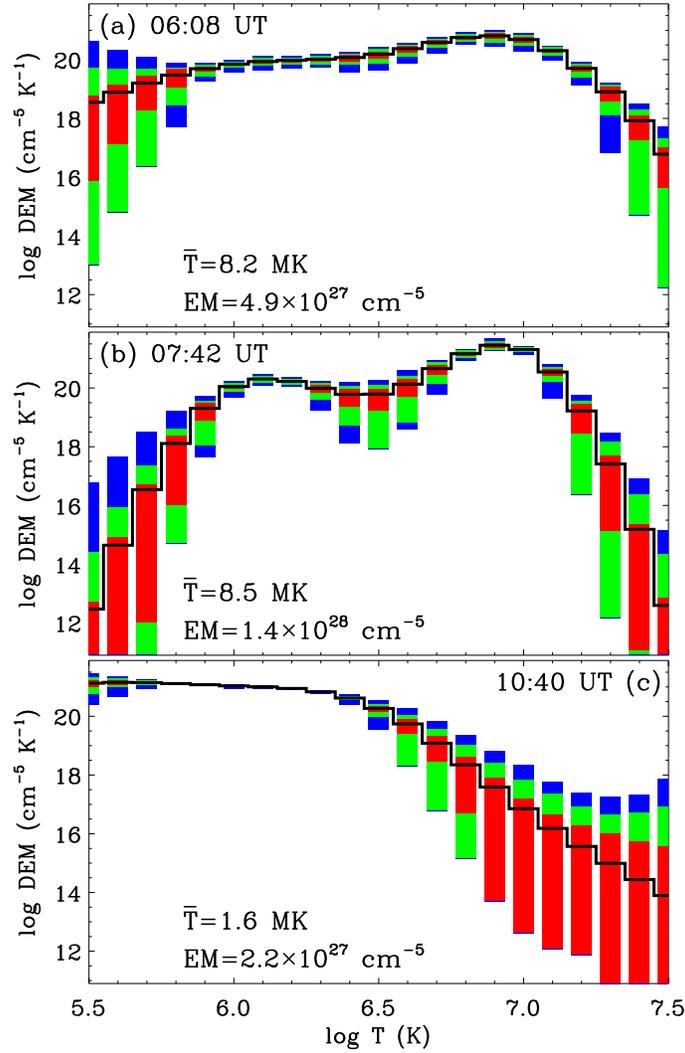}
\caption{Temperature and EM of the quiescent loops.
(a)-(c) DEM curves for the loop regions enclosed by the red rectangles in Figures\,\ref{f:loop_heating}(a)-(b), \ref{f:loop_heating}(c)-(d), and \ref{f:loop_cooling}(e), respectively.
The black lines are the best-fit DEM distributions, and the red, green, and blue rectangles separately represent the regions containing 50\%, 51\%-80\%, and 81\%-95\% of the Monte Carlo solutions. 
See Section\,\ref{sec:res} for details.
\label{f:dems}}
\end{figure}

\begin{figure}[ht!]
\centering
\includegraphics{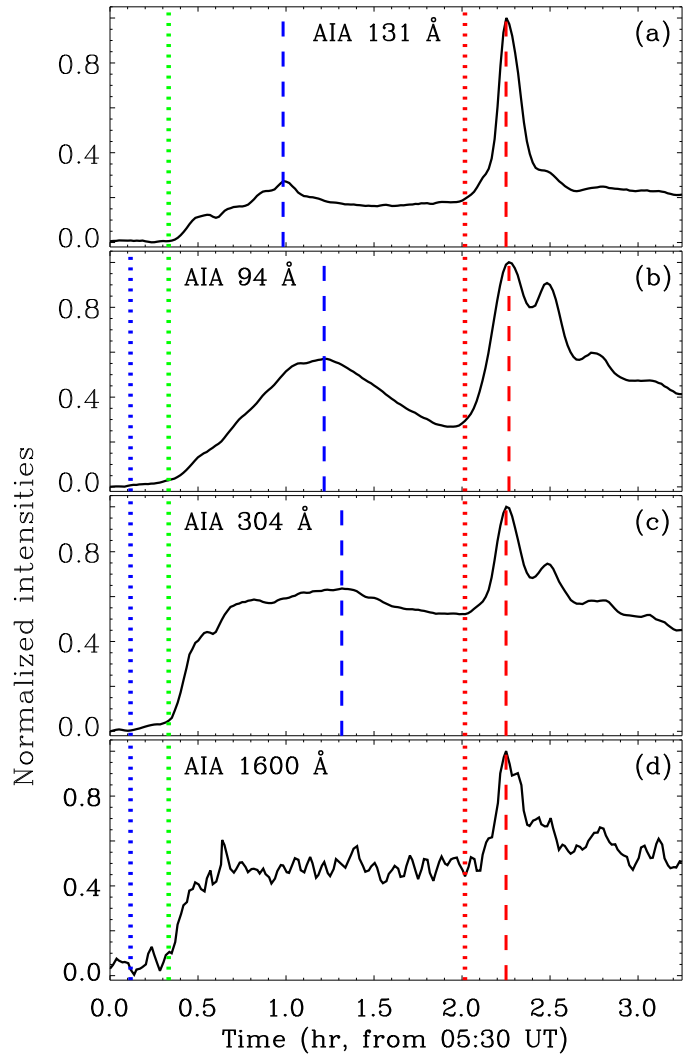}
\caption{Light curves of the heated quiescent loops caused by nearby eruptions.
(a)-(d) Light curves of the AIA 131\,\AA, 94\,\AA, 304\,\AA, and 1600\,\AA~channels in the white and blue rectangles in Figures\,\ref{f:loop_heating} and \ref{f:heating_evolution}, respectively.
The blue, green, and red vertical dotted lines mark the start of heating, and the blue and red vertical dashed lines denote the peaks of light curves.
\label{f:lightcurves1}}
\end{figure}

\begin{figure}[ht!]
\centering
\includegraphics[width=0.8\textwidth]{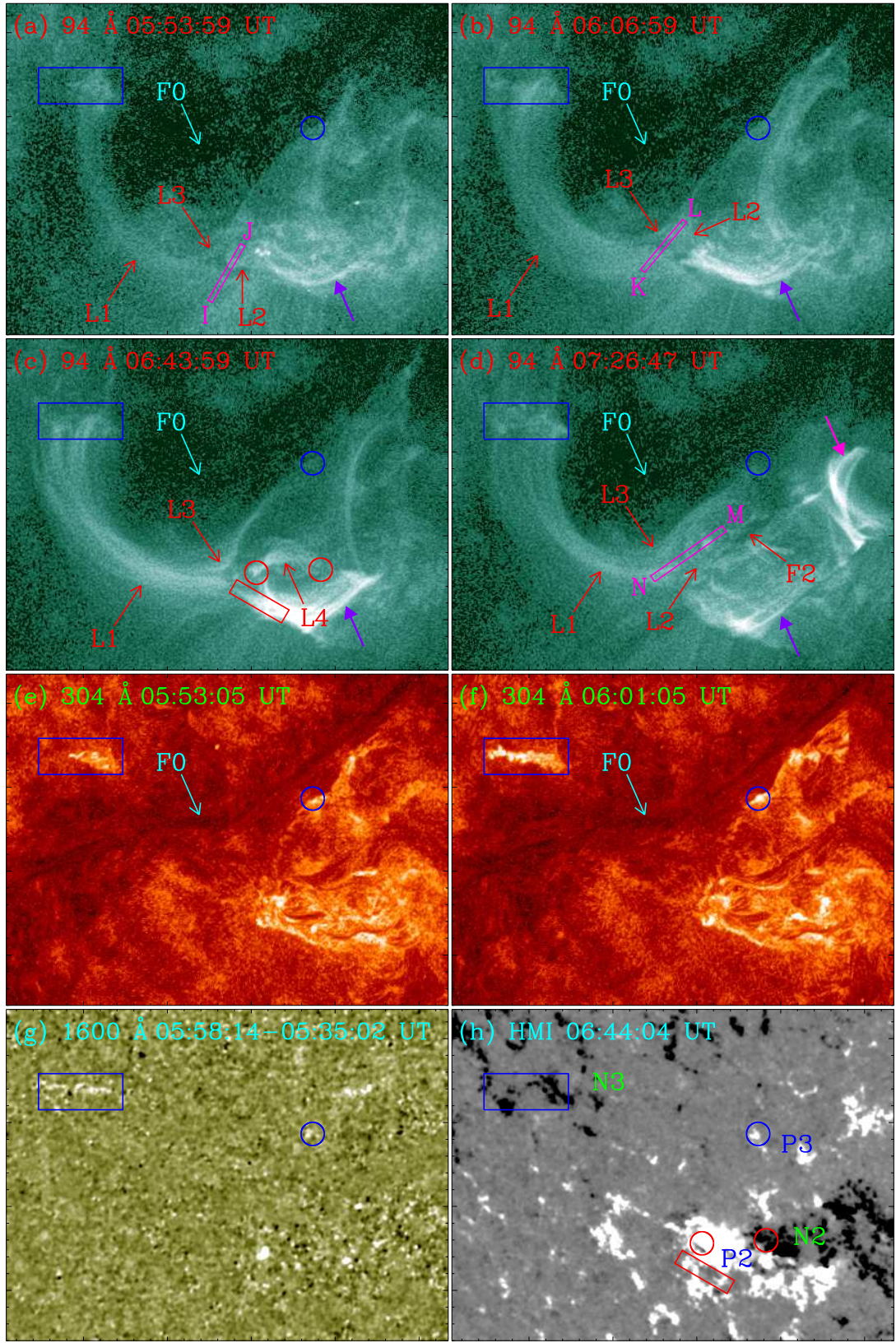}
\caption{Evolution of the heated quiescent loops caused by nearby eruptions.
(a)-(d) AIA 94\,\AA~and (e)-(f) 304\,\AA~images, (g) 1600\,\AA~difference images, and (h) an HMI LOS magnetogram.
The AIA images have been enhanced by the MGN technique.
The blue and red rectangles in (a)-(h) and (c) and (h) mark the northern and southern endpoints of loops L1.
The blue circles in (a)-(h) enclose the western endpoint of loops L3.
The red circles in (c) and (h) denote two endpoints of loops L4.
The purple and pink solid arrows in (a)-(d) indicate the postflare loops caused by the eruptions of warm channel F1 and filament F2, respectively.
The pink rectangles IJ, KL, and MN in (a), (b), and (d) mark the positions for the time slices in Figures\,\ref{f:time_slices2}(a)-(c).
The FOV is indicated by the cyan rectangle in Figure\,
\ref{f:loop_heating}(a).
See Section\,\ref{sec:res} for details.
\label{f:heating_evolution}}
\end{figure}

\begin{figure}[ht!]
\centering
\includegraphics{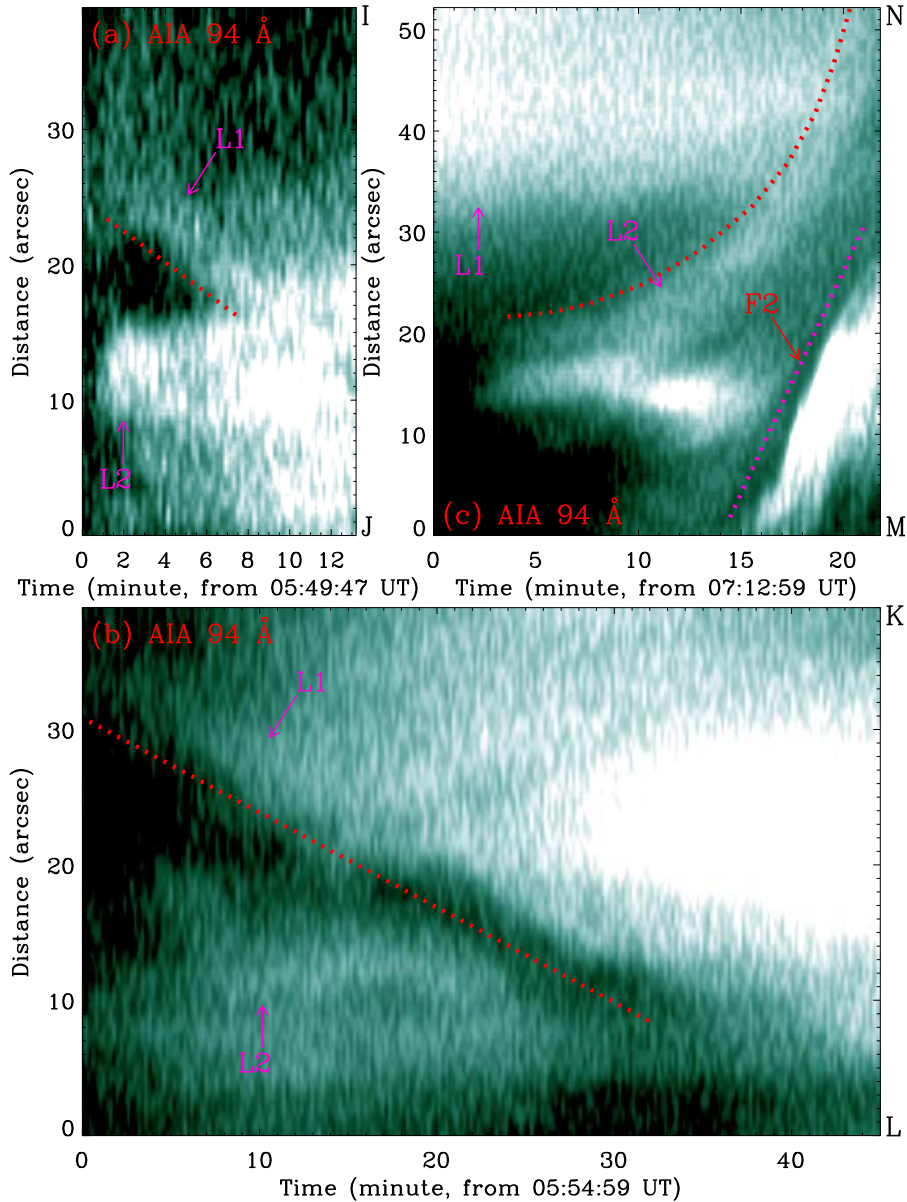}
\caption{Temporal evolution of the quiescent loops.
(a)-(c) Time slices of AIA 94\,\AA~images along the IJ, KL, and MN directions in the pink rectangles in Figures\,\ref{f:heating_evolution}(a), (b), and (d), respectively.
The red dotted lines outline the motion of loops L1 and L2.
The pink dotted line in (c) denotes the eruption of filament F2.
See Section\,\ref{sec:res} for details.
\label{f:time_slices2}}
\end{figure}

\begin{figure}[ht!]
\centering
\includegraphics{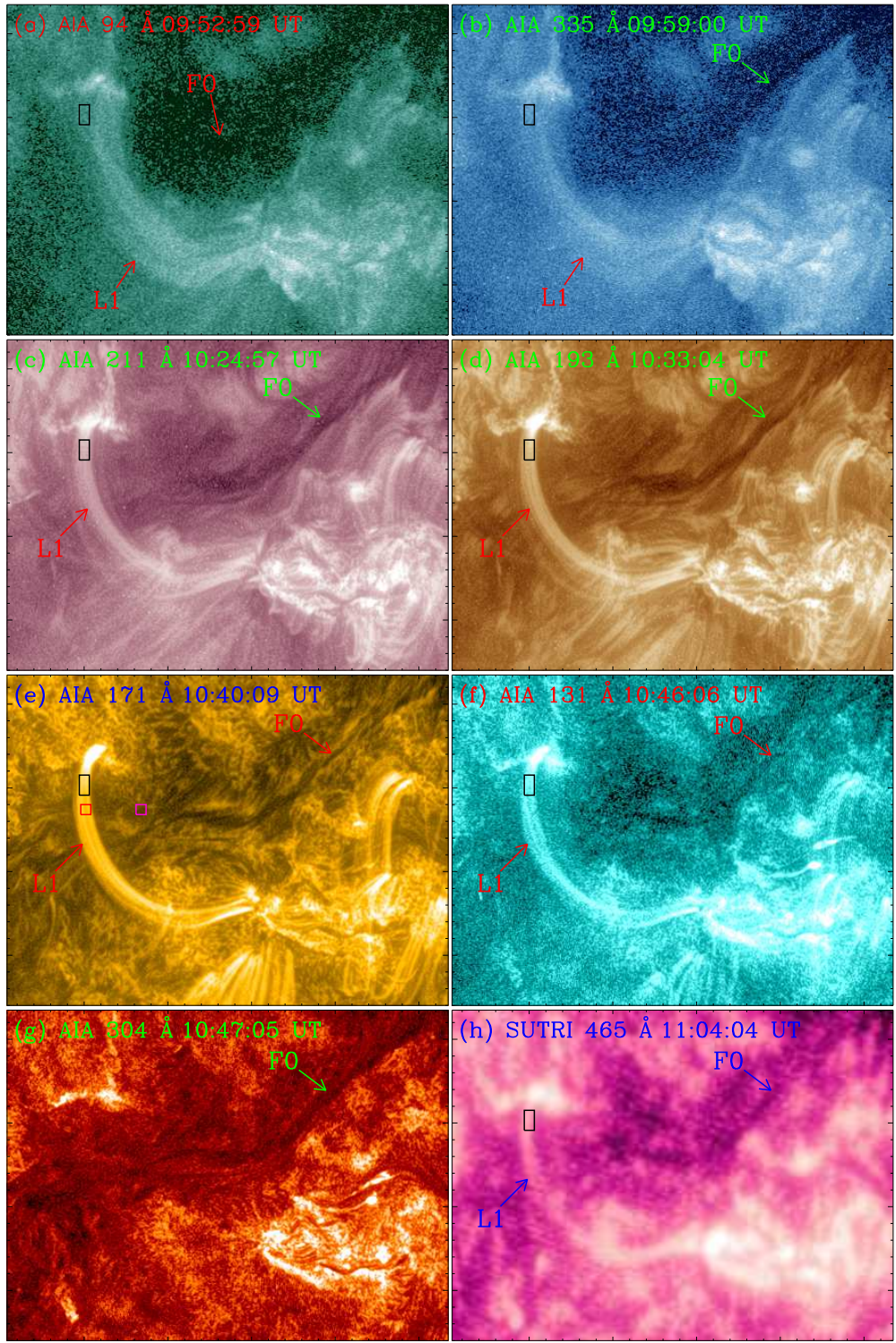}
\caption{Cooling of the heated quiescent loops caused by nearby eruptions.
(a)-(g) AIA 94\,\AA, 335\,\AA, 211\,\AA, 193\,\AA, 171\,\AA, 131\,\AA, 304\,\AA, and (h) SUTRI 465\,\AA~images, enhanced by the MGN technique.
The black rectangles in (a)-(f) and (h) mark the regions for the light curves in Figure\,\ref{f:lightcurves2}.
The red and pink rectangles in (e) enclose the region for the DEM curve in Figure\,\ref{f:dems}(c), and the location where the background emission is computed, respectively.
An animation of the unannotated AIA images (panels (a)-(f)) is available. It covers $\sim$9\,hr starting at 05:00 UT, with a time cadence of 1 minute.
The FOV is indicated by the cyan rectangle in Figure\,\ref{f:loop_heating}(a).
See Section\,\ref{sec:res} for details.
(An animation of this figure is available.)
\label{f:loop_cooling}}
\end{figure}

\begin{figure}[ht!]
\centering
\plotone{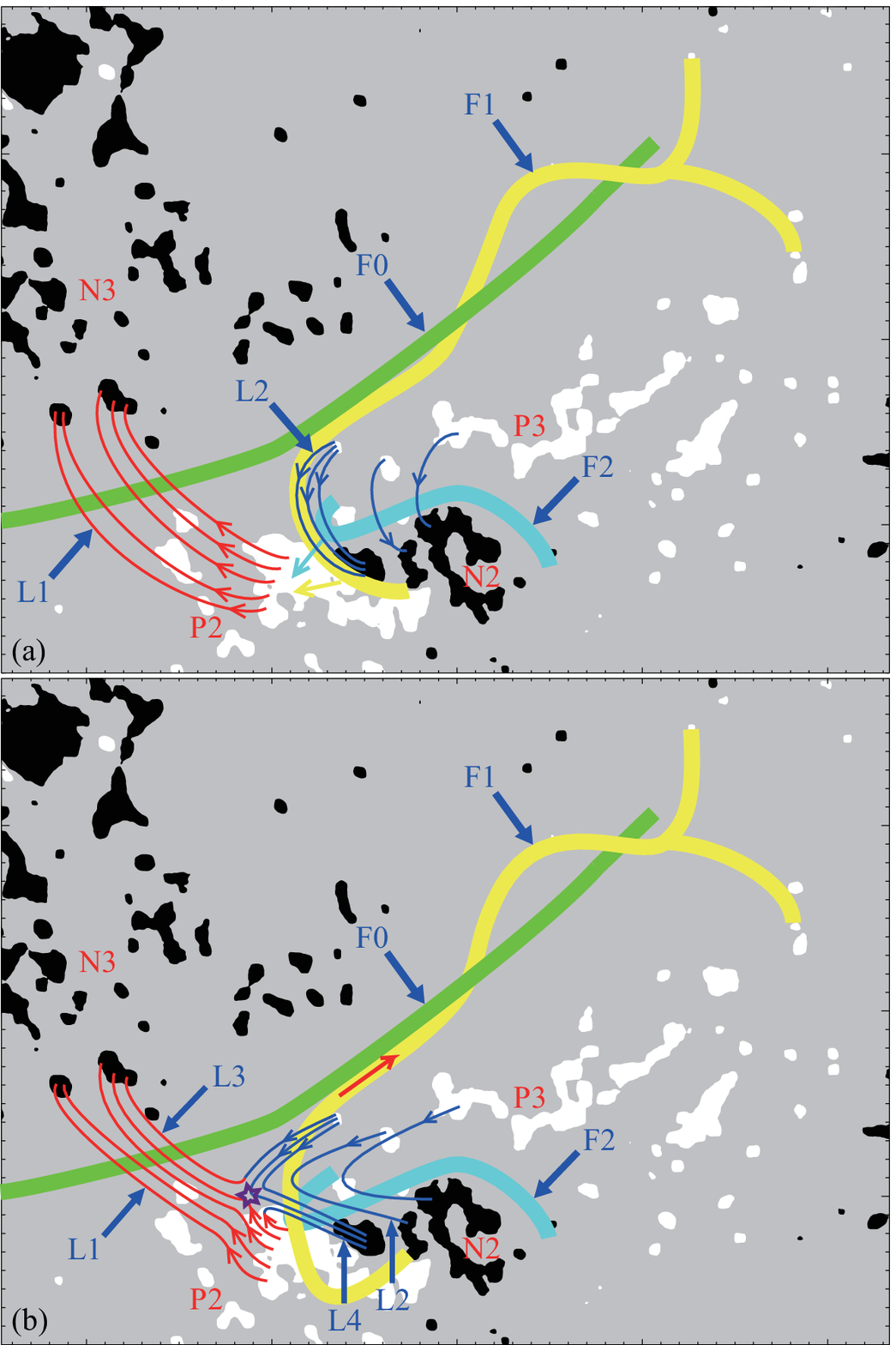}
\caption{Schematic diagrams of the heating of quiescent loops caused by nearby eruptions.
The white and black patches separately indicate the positive and negative magnetic fields P2 and P3, and N2 and N3.
The green, cyan, and yellow thick lines show the filaments F0 and F2, and the warm channel F1.
The red, blue, and red blue lines represent the magnetic field lines of loops L1, L2, and L3 and L4, respectively, whose directions are marked by the red and blue arrows.
The yellow and cyan arrows in (a) indicate the erupting directions of warm channel F1 and filament F2, and the red arrow in (b) shows the propagation of material of the warm channel F1.
The purple star in (b) denotes the magnetic reconnection between loops L1 and L2.
The FOV is indicated by the red rectangle in Figure\,\ref{f:general_information}(a).
See Section\,\ref{sec:sum} for details.
\label{f:cartoon}}
\end{figure}

The negative magnetic fields N2 are encompassed by the positive ones P2 and P3; see Figure\,\ref{f:general_information}(a).
They may separately correspond to the inner parasitic and outer parent fields of a fan-spine topology, which is naturally formed upon these fields at the coronal heights \citep{1990ApJ...350..672L, 1998ApJ...502L.181A}.
This is further supported by the coronal magnetic field obtained from the PFSS model, indicated by the blue lines, connecting the negative fields N2 and the positive ones P2 and P3, and the red and green lines, connecting the positive fields P2 and P3 and the nearby negative ones N3, in Figure\,\ref{f:pfss_fields}.
Here, the outer spine of the fan-spine topology; see the red and green lines, connects to the negative fields N3.
The reconnection between the loops L1 and L2 caused by the nearby eruptions is thus similar to those, which are triggered by the filament eruptions within the fan-spine configuration and result in the circular-ribbon flares, at the coronal null points of fan-spine topology \citep{2017ApJ...851...30X, 2020ApJ...900..158Y}.
Two-ribbon flares, rather than circular-ribbon ones, are observed in this study underneath the eruptions.
This may be caused by the following two reasons: (1) the positions of the erupting structures, i.e., the warm channel (filament) F1 (F2) is located upon the PILs between the positive and negative fields P2 (P3) and N2 rather than the PILs surrounding the inner negative fields N2, and (2) the erupting directions, i.e., the filament (warm channel) F2 (F1) erupted sideward rather than upward toward the coronal null points.
During the reconnection, the newly reconnected loops, e.g., postflare loops, are heated.
In this study, the postflare loops underneath the eruptions; see Section\,\ref{subsec:eruptions}, and the newly reconnected loops L3 and L4; see Section\,\ref{subsec:heating_loops}, are heated, consistent with the previous studies \citep{2009ApJ...690..347L, 2021ApJ...908..213L}. 
Here, the loops L3 and their northern footpoints may separately correspond to the heated outer spine and the remote ribbon during the circular-ribbon flares \citep{2013ApJ...778..139S, 2017ApJ...851...30X}.
Besides the newly reconnected loops, the quiescent loops L1, the inflowing loops toward the reconnection region; see Figures\,\ref{f:heating_evolution} and \ref{f:time_slices2}, are also heated.
They are filled by the evaporated chromospheric material heated by the beamed nonthermal particles and/or thermal conduction front, produced during the reconnection, from the reconnection region to the chromosphere along the newly reconnected loops L3 (L4) that have a similar endpoint to loops L1.

\begin{figure}[ht!]
\centering
\includegraphics{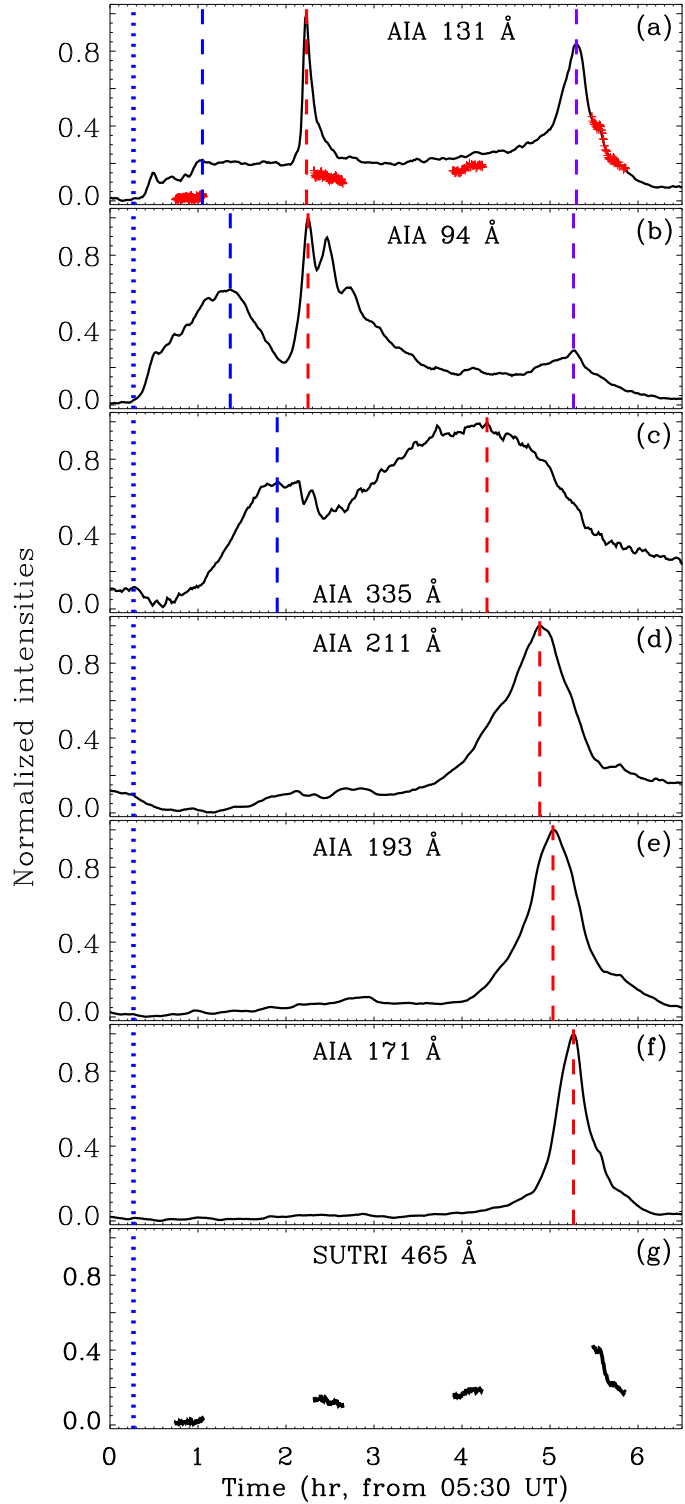}
\caption{Light curves of the cooling quiescent loops.
(a)-(g) Light curves of the AIA 131\,\AA, 94\,\AA, 335\,\AA, 211\,\AA, 193\,\AA, 171\,\AA, and SUTRI 465\,\AA~channels in the black rectangles in Figure\,\ref{f:loop_cooling}.
The blue vertical dotted lines mark the start of heating.
The blue, red, and purple vertical dashed lines show the peaks of light curves.
The red pluses in (a) illustrate the SUTRI 465\,\AA~light curve in (g).
See Section\,\ref{sec:res} for details.
\label{f:lightcurves2}}
\end{figure}

Cooling of the heated quiescent loops caused by nearby eruptions is identified. 
The sequence of the brightening of the loops first in the hot and then in the progressive cooler channels shows the cooling process of the heated loops; see Section\,\ref{subsec:cooling_loops}. 
This is also supported by the AIA and SUTRI EUV light curves; see Figure\,\ref{f:lightcurves2}.
After the eruption of warm channel F1, the cooling loops appeared in AIA 335\,\AA~images, and less of them are detected in AIA 211 and 193\,\AA~images, because they are heated again by reconnection between loops, that is caused by the following eruption of filament F2, before they cooled further down.  
After the eruption of filament F2, the heated loops cooled down from $\sim$7.2\,MK (AIA 94\,\AA) to $\sim$2.5\,MK (AIA 335\,\AA) in $\sim$122\,minutes, to $\sim$1.9\,MK (AIA 211\,\AA) in $\sim$158\,minutes, to $\sim$1.5\,MK (AIA 193\,\AA) in $\sim$167 minutes, and to $\sim$0.9\,MK (AIA 171\,\AA) in $\sim$181 minutes, much longer than those in \citet{2015A&A...583A.109L}.
Moreover, the cooling loops in the SUTRI 465\,\AA~channel, with the characteristic temperature of $\sim$0.5\,MK \citep{2017RAA....17..110T, 2023RAA}, are observed for the first time.
Their evolution is consistent with that of the AIA 131\,\AA~loops; see Figure\,\ref{f:lightcurves2}, which show the plasma with the lower characteristic temperature ($\sim$0.6\,MK) of the AIA 131\,\AA~channel.
In the future, combining the data of the SUTRI, SDO, and Interface Region Imaging Spectrograph \citep[IRIS;][]{2014SoPh..289.2733D}, more heating and cooling events can be investigated, in order to understand the mystery of transition region and coronal heating.

\clearpage

\acknowledgments

The authors thank the referee for helpful comments that led to improvements in the manuscript. We are indebted to the SDO and SUTRI teams for providing the data.
This work is supported by the National Key R\&D Programs of China (2019YFA0405000 and 2021YFA1600502 (2021YFA1600500)),  the National Natural Science Foundations of China (12073042, U2031109, 11825301, and 12073077), the Key Research Program of Frontier Sciences (ZDBS-LY-SLH013) and the Strategic Priority Research Programs (No. XDB 41000000) of CAS, and Yunnan Academician Workstation of Wang Jingxiu (No. 202005AF150025).
AIA images are the courtesy of NASA/SDO and the AIA, EVE, and HMI science teams.
SUTRI is a collaborative project conducted by the National Astronomical Observatories of CAS, Peking University, Tongji University, Xi'an Institute of Optics and Precision Mechanics of CAS, and the Innovation Academy for Microsatellites of CAS.
We acknowledge the usage of Jhelioviewer software \citep{2017A&A...606A..10M}, and NASA's Astrophysics Data System.


\end{document}